\documentclass[floatfix,%
reprint,
showpacs,preprintnumbers,
amsmath,amssymb,
prb,
superscriptaddress
]{revtex4-2}

\usepackage{graphicx}
\usepackage{dcolumn}
\usepackage{bm}
\usepackage{physics}
\usepackage{comment}
\usepackage{algpseudocode}
\graphicspath{{./figures/}}
\usepackage{color}

\newcommand{\NEM}{Department of Nuclear Engineering and Management, Graduate School of Engineering, The University of Tokyo, 7-3-1 Hongo, Bunkyo-ku, Tokyo 113-8656, Japan}
\newcommand{\PSC}{Photon Science Center, Graduate School of Engineering, The University of Tokyo, 7-3-1 Hongo, Bunkyo-ku, Tokyo 113-8656, Japan}
\newcommand{\UTripl}{Research Institute for Photon Science and Laser Technology, The University of Tokyo, 7-3-1 Hongo, Bunkyo-ku, Tokyo 113-0033, Japan}
\newcommand{\ALFA}{Institute for Attosecond Laser Facility, The University of Tokyo, 7-3-1 Hongo, Bunkyo-ku, Tokyo 113-0033, Japan}
\newcommand{\KPSI}{Kansai Institute for Photon Science, National Institutes for Quantum Science and Technology, Umemidai, Kizugawa-shi, Kyoto 619-0215, Japan}
\newcommand{\AIST}{Materials DX Research Center, National Institute of Advanced Industrial Science and Technology, Umezono, Tsukuba, Ibaraki 305-8568, Japan}

\begin{document}
\title{Ultrafast Energy Absorption in Silicon Controlled by Two-Color Double Pulses}

\author{Eiyu S. Gushiken}
\email{gushiken-eiyu20003@g.ecc.u-tokyo.ac.jp}
\affiliation{\NEM}

\author{Mizuki Tani}
\affiliation{\NEM}
\affiliation{\KPSI}

\author{Hiroki Katow}
\altaffiliation{Present address: \AIST}
\affiliation{\PSC}

\author{Kenichi L. Ishikawa}
\email[corresponding author:]{ishiken@n.t.u-tokyo.ac.jp}
\affiliation{\NEM}
\affiliation{\PSC}
\affiliation{\UTripl}
\affiliation{\ALFA}


\date{\today}

\begin{abstract}
We theoretically show that energy absorption in crystalline silicon can be controlled by two-color femtosecond double-pulse irradiation, in which two temporally separated pulses with different wavelengths interact sequentially with the system.
Using time-dependent density functional theory, we systematically examine the wavelength and intensity dependence of the absorbed energy over peak intensities of $2\times10^{11}$--$10^{13}$ W/cm$^2$ and wavelengths of 515, 1030, and 2060 nm.
We find that the mechanism governing energy absorption and the optimal wavelength combination strongly depend on the intensity regime. 
In the low-intensity regime, multiphoton interband absorption is dominant, and energy absorption is enhanced for pulse pairs composed of shorter wavelengths. 
In contrast, in the high-intensity regime, the contributions of tunneling ionization and intraband acceleration become significant, leading to enhanced absorption for longer-wavelength combinations. 
In the intermediate-intensity regime, a pronounced enhancement is observed when a short-wavelength pulse precedes a long-wavelength pulse.
Our analysis reveals that the nonequilibrium electronic state prepared by the first pulse modifies the excitation process induced by the second pulse, thereby enhancing the absorbed energy through an increased energy gain per excited electron. 
In this regime, the energy absorption is governed not only by the number of excited carriers but also by the energy gain per excited electron, which can be strongly modified by the pulse sequence.
These results indicate that ultrafast energy transfer in semiconductors is tunable by appropriately designing the wavelength and intensity combination of the two pulses, and provide microscopic insight into two-color strong-field excitation.
\end{abstract}

\pacs{Valid PACS appear here}
\maketitle

\section{\label{sec:introduction}INTRODUCTION}
The interaction of ultrashort intense laser pulses, typically operating in the femtosecond to picosecond regime, with solid materials has attracted considerable attention not only for scientific interest, such as high-order harmonic generation~\cite{Ghimire11, Schubert14, Hohenleutner15, Luu15}, but also for medical and industrial applications~\cite{Sugioka14, Gattass08, Malinauskas16, Yang25, Ma25}.
In particular, ultrashort pulse lasers enable nonthermal ablation of solid materials by depositing energy into the electronic subsystem faster than the lattice can respond thermally~\cite{Chichkov96, Gattass08, Tinten98, Stuart96, Kerse16, Liu97}.
This ultrafast energy transfer enables highly localized, high-precision material modification with minimal thermal damage, making ultrashort pulse lasers reliable tools for industrial, biomedical, and scientific applications, including microelectronics, photonics, and medical device fabrication, where nanometer-scale precision is often required~\cite{Sugioka14}.

Among the diverse ultrashort laser processing techniques, burst-mode irradiation, where multiple pulses are delivered in rapid succession, has attracted significant attention for its potential to improve ablation efficiency, particularly under optimized conditions~\cite{Kerse16, Mishchik19, Bonamis19, Francesc22, Zubauskas25}.
The reported enhancement is often associated with mechanisms such as ablation cooling, in which residual heat from preceding pulses is partially dissipated before subsequent pulses arrive, thereby promoting efficient material removal with limited thermal damage.
In parallel, double-pulse irradiation, typically involving two pulses with a controlled time delay, provides a complementary approach~\cite{Kevin20, Zukerstein24, Yu14, Gedvilas17, Chu20, Sergey21, Zhang19, Zoppel07, Guillaume22}.
One notable example of such laser processing is the Transient and Selective Laser (TSL) method developed by Ito \textit{et al.}, where an ultrashort pulse generates a free-electron region in a transparent material, followed by selective absorption of a low-intensity, long-duration pulse into the modified region, enabling rapid and precise micromachining~\cite{Ito2018, Yoshizaki21, Zhang25}.
In both burst-mode and double-pulse processing, the effect of relaxation, lattice motion, and thermal diffusion between pulses introduces complex material responses. 
However, when the temporal separation between the pulses approaches or becomes shorter than the characteristic timescales of electronic decoherence, which are typically several tens of femtoseconds, coherent electron dynamics can dominate the energy absorption process~\cite{Schultze14, Wachter14}.
In such regimes where coherent electron dynamics play a dominant role, it remains an open question how the energy absorption process in semiconductors like crystalline silicon responds to variations in double-pulse parameters.

Furthermore, the use of synthesized dual-color laser pulse pairs is reported to enable highly efficient laser ablation of transparent materials, compared to single-color irradiation \cite{Sugioka93, Zhang97, Zhang00, Obata01, Zoppel05, Zoppel07, Yu13, Yu14, Yang15, Gedvilas17, Tani22}.
For instance, we have recently utilized the time-dependent density functional theory (TDDFT) ~\cite{Runge84} to investigate electron dynamics in silicon under two-color, temporally overlapping laser fields~\cite{Tani22}.
This work has revealed that the temporal overlap of the pulses significantly enhances energy absorption.
This enhancement has been attributed to the interplay between infrared (IR)-driven intraband motion and ultraviolet (UV)-induced interband transitions, with the total energy deposition being maximized when the two pulse energies are approximately equal.

In the present study, we investigate energy absorption in bulk silicon under two temporally separated femtosecond laser pulses with different wavelengths using TDDFT.
The TDDFT~\cite{Runge84} is a first-principles method widely used to describe nonequilibrium electron dynamics in materials under intense laser fields~\cite{Tani21, Kozák20, Miyamoto21, Otobe16, Nicolas17, Tancogne18, Nicolas18, Floss19, Yamada21}.
It captures non-adiabatic electron dynamics, including multiphoton excitation, dielectric breakdown, high-harmonic generation, and transient charge redistribution~\cite{Otobe08, Chu12, Sato14N, Schultze14, Sato14F, Tancogne17, Lian18, Yu19, Yamada19M, Sato25}.
A rigorous first-principles description of actual double-pulse laser processing is challenging due to its multiscale and multiphase nature.
Instead, we focus on the initial transient, largely coherent dynamics induced by femtosecond double pulses, before electron--phonon energy transfer sets in (typically on timescales of $\gtrsim 200\,\mathrm{fs}$) and local thermodynamic equilibrium is established. 
We employ the SALMON code~\cite{Noda19} and examine the dependence of energy absorption on the wavelength and intensity combinations of the two pulses. 
Our analysis shows that the nonequilibrium electronic state prepared by the first pulse influences the response to the second pulse, resulting in a strong dependence of the absorbed energy on the pulse parameters and a marked enhancement under specific double-pulse conditions. 
These results indicate that ultrafast energy transfer in silicon is tunable through the wavelength and intensity pairing of the two pulses.

This paper is organized as follows. 
Section~\ref{sec:TDDFT} describes our simulation methods. We present the numerical results and discussion in Sec.~\ref{sec:results}. 
Conclusions are given in Sec.~\ref{sec:conclusion}.

\section{\label{sec:TDDFT}Time-dependent density functional theory}

The temporal evolution of an $N$-electron system subjected to an optical field is calculated by solving the time-dependent Kohn–Sham (TDKS) equations for the Kohn–Sham orbitals $\psi_{b,\mathbf{k}}(\mathbf{r},t)$~\cite{Runge84}.
Here, $b$ and $\mathbf{k}$ denote the band index and the crystal momentum, respectively.
Spin indices are omitted.
The TDKS equations read,
\begin{equation}
    i\hbar \frac{\partial}{\partial t}\psi_{b,\mathbf{k}}(\mathbf{r},t) = H_{\mathrm{KS}}[n(\mathbf{r},t)]\psi_{b,\mathbf{k}}(\mathbf{r},t),
    \label{TDKS}
\end{equation}
where
\begin{equation}
    H_{\mathrm{KS}}[n(\mathbf{r},t)] = \frac{1}{2m_e}\left[ \mathbf{p}+\dfrac{e}{c}\mathbf{A}(t) \right]^2 + V_{\mathrm{eff}}[n(\mathbf{r},t)],
\end{equation}
denotes the Kohn-Sham Hamiltonian in the velocity gauge within the electric dipole approximation, $\mathbf{p}$ is the canonical momentum, $e$ is the elementary charge, $c$ is the speed of light, $\mathbf{A}(t)$ is the vector potential, and $m_e$ is the electron mass.
The time-dependent electron density is calculated from the occupied orbitals as
\begin{equation}
    n(\mathbf{r},t) = 2 \sum_{\mathbf{k}} \sum_{b\in \mathrm{occ}}
    |\psi_{b,\mathbf{k}}(\mathbf{r},t)|^2.
\end{equation}
The prefactor of 2 accounts for the spin degeneracy of the orbitals, which is assumed in this work.
The effective potential $V_{\mathrm{eff}}$ is composed of three contributions:
\begin{equation}
    V_{\mathrm{eff}}[n(\mathbf{r},t)] = V_{\mathrm{ion}}(\mathbf{r}) + V_{\mathrm{H}}[n(\mathbf{r},t)] + V_{\mathrm{xc}}[n(\mathbf{r},t)],
    \label{Veff}
\end{equation}
where $V_{\mathrm{ion}}$ is the electron–ion interaction represented by norm-conserving pseudopotentials \cite{Fuchs99}, $V_{\mathrm{H}}$ is the Hartree potential, and $V_{\mathrm{xc}}$ is the exchange–correlation potential. In this work, we use the modified Becke–Johnson potential for $V_{\mathrm{xc}}$ \cite{Tran09, Sato15}.
In the velocity gauge, the lattice periodicity of the Kohn--Sham
Hamiltonian is preserved, and the orbitals can be expressed in Bloch
form as
\begin{equation}
    \psi_{b,\mathbf{k}}(\mathbf{r},t)
    = e^{i\mathbf{k}\cdot\mathbf{r}} u_{b,\mathbf{k}}(\mathbf{r},t),
\end{equation}
where $u_{b,\mathbf{k}}(\mathbf{r},t)$ denotes the lattice-periodic part of the Bloch function.

We consider a two-color double-pulse laser field described by the vector potential
\begin{align}
    \mathbf{A}(t) &= \mathbf{A}_1(t)+\mathbf{A}_2(t), \label{At} \\
    \mathbf{A}_1(t) &= -\mathbf{a_1}\cos^2{\left[ \frac{\pi}{T}\left( t-\frac{T}{2} \right) \right]}\sin{\left[ \omega_1\left( t-\frac{T}{2} \right) \right]} \notag \\
    &\qquad\qquad\qquad\qquad\qquad\qquad\qquad (0\le t\le T), \\
    \mathbf{A}_2(t) &=  -\mathbf{a_2}\cos^2{\left[ \frac{\pi}{T}\left( t-\frac{T}{2}-T_{\mathrm{delay}} \right) \right]} \notag \\
    &\qquad\qquad\quad \times\sin{\left[ \omega_2\left( t-\frac{T}{2}-T_{\mathrm{delay}} \right) \right]} \notag \\
    &\qquad\qquad\qquad\qquad\quad (T_{\mathrm{delay}}\le t\le T_{\mathrm{delay}}+T),
\end{align}
where $\mathbf{a}_1$ and $\mathbf{a}_2$ are the amplitude and polarization vectors of each pulse, $T$ is the foot-to-foot pulse duration, for which the full-width-at-half-maximum (FWHM) pulse duration is $0.36T$, $\omega_1$ and $\omega_2$ are the central frequencies of each pulse, and $T_{\mathrm{delay}}$ is the time delay between the pulses. 
Outside their respective time domain of definition, $\mathbf{A}_1$ and $\mathbf{A}_2$ are set to zero. 

The absorbed energy is evaluated as the work done by the electric field $\mathbf{E}(t) = -\dfrac{1}{c}\dot{\mathbf{A}}(t)$,
\begin{equation}
    \mathrm{W(}t) = \int_0^t d\tau \, \mathbf{J}(\tau) \cdot \mathbf{E}(\tau),
\end{equation}
where the current density $\mathbf{J}(t)$ is given by
\begin{equation}
    \mathbf{J}(t) = -i\frac{2e}{\hbar \Omega} \int_{\Omega} d\mathbf{r} \sum_{\mathbf{k}}\sum_{b} \psi_{b,\mathbf{k}}^*(\mathbf{r}, t) \left[ H_{\mathrm{KS}}, \mathbf{r} \right] \psi_{b,\mathbf{k}}(\mathbf{r}, t),
    \label{Jt}
\end{equation}
with $\Omega$ being the volume of the simulation box.
The calculations are performed on a simple cubic unit cell containing eight Si atoms in the diamond structure, with a lattice constant of 5.43~\AA~\cite{Massa09}. 
The real space and the reciprocal space are each discretized into $24^3$ grids.
Periodic boundary conditions are applied in all the three directions.
Details of the ground-state electronic structure and linear-response properties are given in Appendix~\ref{sec:appendix}.

\section{\label{sec:results}RESULTS AND DISCUSSION}

\begin{table}[tb]
  \centering
  \caption{Summary of simulation parameters used throughout this study.}
  \begin{tabular}{ll}
    \hline
    \textbf{Parameter} & \textbf{Value} \\
    \hline
    Time step & 0.75~as\\
    Total simulation time & 65~fs \\
    Pulse width (FWHM) & 10.8~fs \\
    Wavelengths $\lambda$ & 515, 1030, 2060~nm \\
    (Photon energy $\hbar\omega$) & $\left( 2.41,~1.20,~0.602~\rm{eV} \right)$ \\
    Peak intensity $I$ & $2 \times 10^{11}$ – $10^{13}$~W/cm$^2$ \\
    Polarization direction & [001] \\
    \hline
  \end{tabular}
  \label{tab:parameters}
\end{table}

We examine how the absorbed energy varies with the combination of the wavelength $\lambda_1$ and $\lambda_2$ with the peak intensity of the two pulses being the same $I_1 = I_2$. 
The simulation parameters are summarized in Table~\ref{tab:parameters}.
In this study, the combination of the first pulse at wavelength $\lambda_1$ and the second pulse at $\lambda_2$ is denoted as “$\lambda_1 + \lambda_2$ nm”.

\subsection{\label{subsec:moderate}Wavelength-combination dependence at moderate intensity}

\begin{figure}[tb]
  \centering
  \includegraphics[width=1\linewidth]{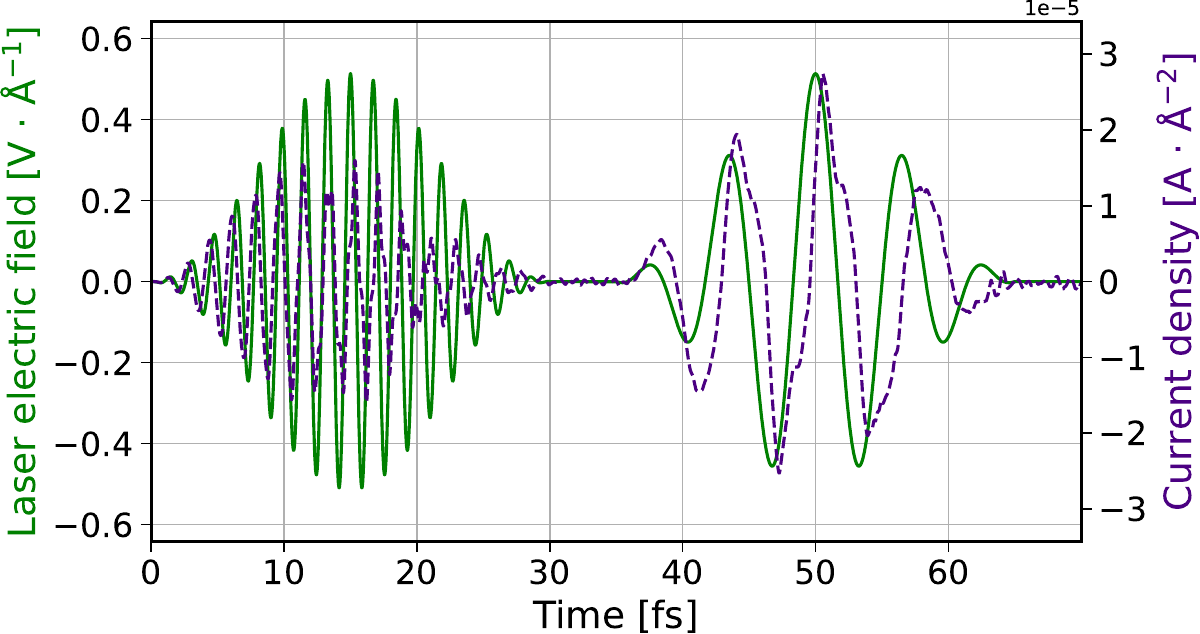}
  \caption{
    Temporal profile of the electric field $\mathrm{E}(t)$ (green solid, left axis) and induced current density $\mathrm{J}(t)$ (purple dashed, right axis) for a double-pulse 515 + 2060 nm, $I_1 = I_2 = 3.5 \times 10^{12}$~W/cm$^2$, $T_{\mathrm{delay}} = 35$~fs.
  }
  \label{fig:E_J_eg}
\end{figure}

In this subsection, we investigate a representative case to understand the fundamental dynamics of energy transfer under double-pulse irradiation at moderate peak intensities ($I_1 = I_2 = 3.5 \times 10^{12}$~W/cm$^2$) and time delay between pulses of $T_{\mathrm{delay}} = 35$ fs.
Figure~\ref{fig:E_J_eg} presents the temporal profile of the electric field $\mathrm{E}(t)$ and the induced current density $\mathrm{J}(t)$ for 515 + 2060 nm double-pulse.
Each pulse has an FWHM pulse duration of 10.8 fs ($T=30\,\mathrm{fs}$), and the two pulses do not overlap with each other.

\subsubsection{\label{subsubsec:moderate_energy_absorption}Energy transfer to electrons}\begin{figure}[tb]
  \centering
  \includegraphics[width=0.9\linewidth]{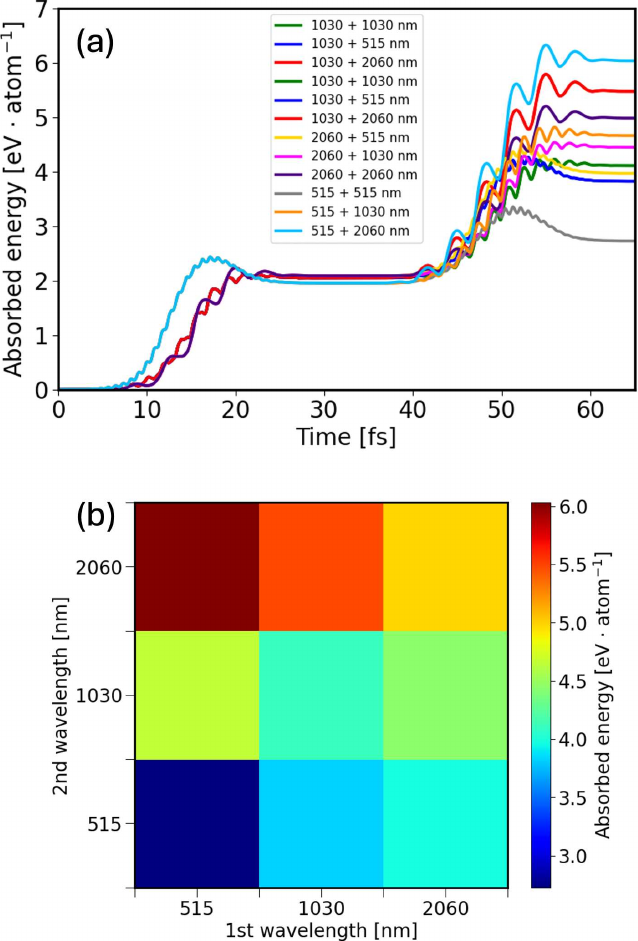}
  \caption{
    (a) Temporal profile of the absorbed energy for each wavelength combination with $I_1 = I_2 = 3.5 \times 10^{12}$ W/cm$^2$ and $T_{\mathrm{delay}} = 35$~fs. 
    (b) Total amount of absorbed energy at the end of the simulation for each combination.
  }
  \label{fig:abs_mid_double}
\end{figure}

We now examine how the energy transfer depends on the combination of wavelengths under the fixed condition of $I_1 = I_2 = 3.5 \times 10^{12}$~W/cm$^2$ and $T_{\mathrm{delay}} = 35$~fs. 
Figure~\ref{fig:abs_mid_double}(a) shows the temporal evolution of the absorbed energy.
The energy deposited by the first pulse (around 30~fs) is found to be nearly independent of its wavelength.
In contrast, the final absorbed energy (around 65~fs) shows a strong dependence on the combination of both pulses.
Notably, the energy deposited by the second pulse depends not only on its wavelength but also on that of the first pulse, indicating the effect of the electronic excitation state prepared by the first pulse.
Figure~\ref{fig:abs_mid_double}(b) shows the final absorbed energy for all combinations of wavelengths 515~nm, 1030~nm, and 2060~nm.
The combination that results in the greatest energy transfer is 515 + 2060 nm.
A general trend is that a longer-wavelength pulse preceded by a shorter-wavelength pulse is advantageous for energy absorption.

\subsubsection{\label{subsubsec:moderate_carriers}Number of excited electrons and mean absorbed energy}
The absorbed energy can be decomposed into two contributions: the number of generated carriers (electrons promoted to the conduction band) and their mean energy absorption.  
Then, is the difference in energy deposition observed in the previous subsection primarily due to that in the former or the latter?
To answer this question, we investigate the dependence of these two quantities on the wavelength combination.  

The number of excited electrons is evaluated by projecting the time-dependent Kohn--Sham orbitals onto the eigenstates of the ground-state Hamiltonian with crystal momentum shifted by the time-dependent vector potential, i.e., the Houston basis.
We first define the electron occupation in the Houston basis as
\begin{equation}
    n_{b,\mathbf{k}}(t)=2\sum_{j}
    \left|
    \left\langle
    \psi^{(0)}_{b,\mathbf{k}+ \frac{e}{\hbar c} \mathbf{A}(t)}
    \middle|
    \psi_{j,\mathbf{k}}(t)
    \right\rangle
    \right|^2 ,
    \label{eq:houston_occ}
\end{equation}
where $\psi^{(0)}_{b,\mathbf{k}+ \frac{e}{\hbar c} \mathbf{A}(t)}$is the ground-state eigenstate evaluated at the shifted crystal momentum $\mathbf{k}+ \frac{e}{\hbar c} \mathbf{A}(t)$.
The total number of excited electrons is then given by
\begin{equation}
    n_{\mathrm{ex}}(t)=\frac{1}{N_k} \sum_{\mathbf{k}}\sum_{b\in\mathrm{cond}} n_{b,\mathbf{k}}(t),
    \label{eq:nex_houston}
\end{equation}
where $b\in\mathrm{cond}$ denotes the conduction-band states and $N_k$ is the total number of $k$-points.

The mean absorbed energy per excited electron is given by
\begin{equation}
    {\mathrm{W_{mean}}}(t)=\frac{\mathrm{W}(t)}{n_\mathrm{{ex}}(t)}.
\end{equation}

 \begin{figure}[tb]
  \centering
  \includegraphics[width=0.9\linewidth]{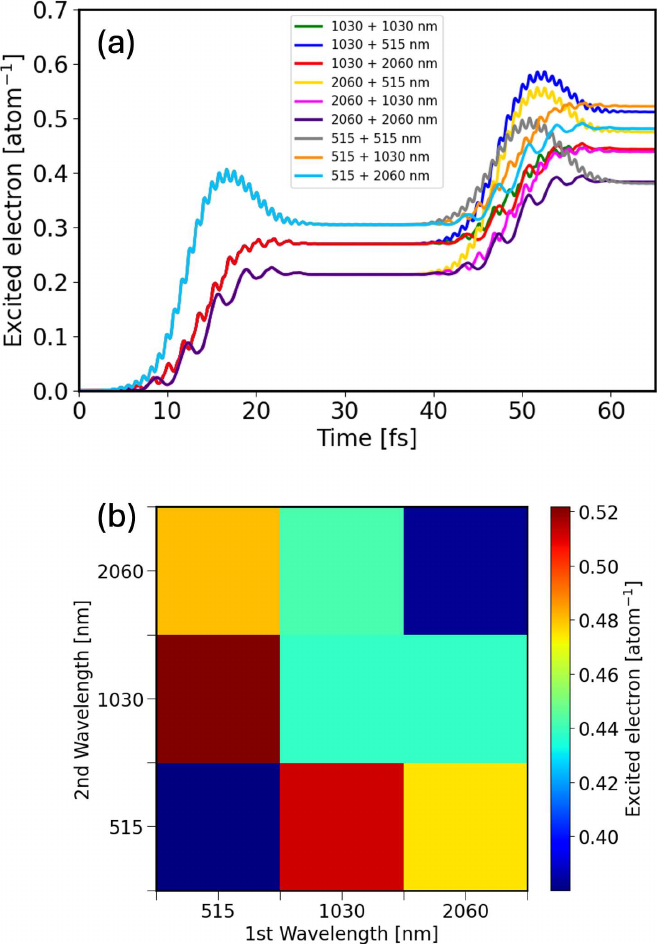}
  \caption{(a) Temporal profile of the calculated number of excited carriers for each wavelength combination with $I_1=I_2=3.5 \times 10^{12}$ W/$\mathrm{cm}^2$ and $T_{\mathrm{delay}}=35$ fs. 
  (b) Total number of excited carriers for each wavelength combination.}
  \label{elc_mid_double}
\end{figure}

Figure~\ref{elc_mid_double} shows the time evolution [panel (a)] and final values [panel (b)] of the number of excited electrons under each wavelength combination.  
When the wavelength is shorter, i.e., the photon energy is higher, the first pulse excites more electrons due to more efficient interband transitions (See around 0--30~fs in panel (a)).
However, the final number of excited electrons after the second pulse does not strongly depend on the wavelength combination, compared with the absorbed energy shown in Fig.~\ref{fig:abs_mid_double}(a).
These results suggest that the increased energy deposition observed for certain double-pulse configurations arises primarily from a higher energy gain per excited carrier, rather than from a larger number of carriers.

\begin{figure}[tb]
  \centering
  \includegraphics[width=0.9\linewidth]{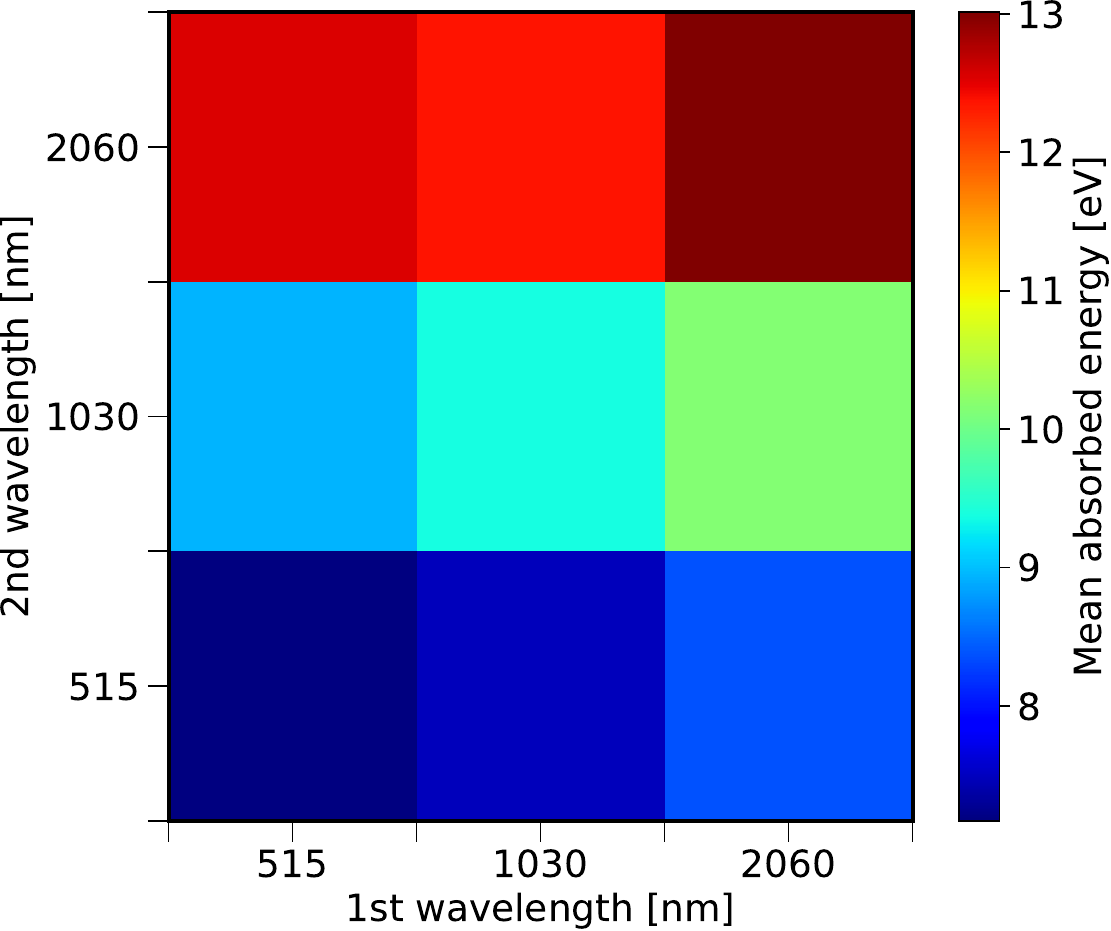}
  \caption{Mean absorbed energy per excited electron for each wavelength combination with $I_1=I_2=3.5 \times 10^{12}$ W/cm$^2$ and $T_{\mathrm{delay}}=35$~fs.}
  \label{mean_mid_double}
\end{figure}

To further verify this speculation, we directly present the mean absorbed energy per electron in Fig.~\ref{mean_mid_double}.  
It clearly shows that configurations in which the second pulse has a longer wavelength result in higher mean energy absorption.

\subsubsection{\label{subsubsec:moderate_ddos}Excitation dynamics analysis}
The findings in the previous subsection motivate a more detailed analysis of the time- and energy-resolved excitation behavior.
To this end, we compute the difference density of states (DDOS), which reflects changes in electron occupancy relative to the ground state.
DDOS is calculated by projecting the time-dependent Kohn-Sham orbitals onto the initial (ground-state) eigenstates.
The DDOS, denoted as $\Delta f(t,\epsilon)$, is then defined as
\begin{equation}
    \Delta f(t,\epsilon) = \frac{1}{N_k} \sum_{\mathbf{k}}\sum_{b} \left[ n_{b,\mathbf{k}}(t) - n_{b,\mathbf{k}}(0) \right] \delta(\epsilon - E_{b,\mathbf{k}}),
\end{equation}
where $E_{b,\mathbf{k}}$ denotes the ground-state Kohn-Sham eigenenergy, and $\delta$ is the Dirac delta function.  
In practical implementation, the delta function is replaced by a Gaussian broadening to yield a smooth spectrum.
At the ground state ($t=0$), electrons occupy only the valence band, with no electrons in the conduction band.
After excitation by a laser pulse, $\Delta f(t,\epsilon)<0$ in the valence band and $\Delta f(t,\epsilon)>0$ in the conduction band.
The absolute values correspond to the number of holes and excited electrons, respectively.
From the conservation of the total number of electrons, we obtain
\begin{equation}
    \int_{-\infty}^{\infty} d\epsilon~\Delta f(t,\epsilon) = 0.
\end{equation}
In practical numerical computations, the projection is carried out using a finite number of ground-state orbitals, and the energy integration is therefore restricted to an energy window $[\epsilon_\mathrm{min},\epsilon_\mathrm{max}]$.
We define the residual charge imbalance within this projection window as
\begin{equation}
    g(t) = \int_{\epsilon_\mathrm{min}}^{\epsilon_\mathrm{max}}
    d\epsilon~\Delta f(t,\epsilon).
\end{equation}
The lower bound $\epsilon_\mathrm{min}$ is chosen sufficiently low to include all initially occupied valence states, while the upper bound is set to $\epsilon_\mathrm{max}=15$~eV.
Since TDDFT conserves the total electron number, any deviation $g(t)\neq 0$ reflects only the finite projection range.
When population is transferred to states above $\epsilon_\mathrm{max}$, the hole count within the window exceeds the excited-electron count, yielding $g(t)<0$.
In this case, $-g(t)$ estimates the number of electrons excited to energies above $\epsilon_\mathrm{max}$.

\begin{figure}[tb]
  \centering
  \includegraphics[width=0.95\linewidth]{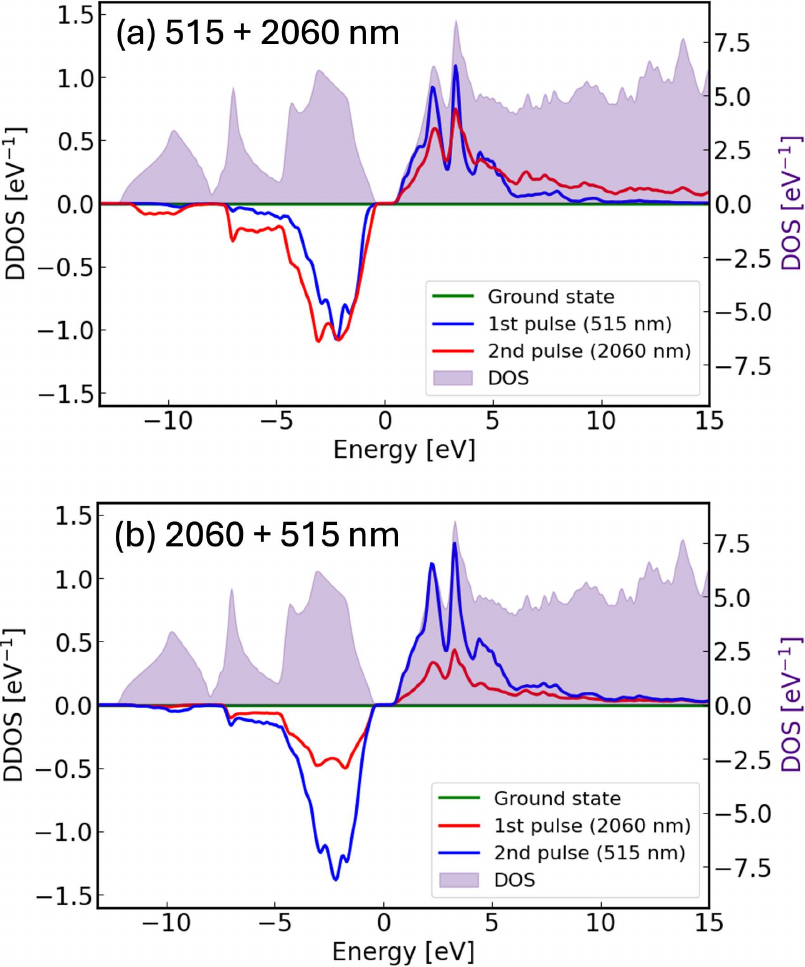}
  \caption{
  DDOS at $t=35$ and $65$~fs, corresponding to the times immediately after the first and second laser pulses, respectively, for $I_1=I_2=3.5\times10^{12}$~W/cm$^2$.
  Panels (a) and (b) show the results for the 515+2060~nm and 2060+515~nm, respectively.
    }
  \label{fig:ddos_mid_double}
\end{figure}

Figure~\ref{fig:ddos_mid_double} shows snapshots of the DDOS at $t=35$ and $65$~fs, corresponding to the times immediately after the first and second laser pulses, respectively, for $I_1=I_2=3.5\times10^{12}$~W/cm$^2$, $T_{\mathrm{delay}}=35$~fs, for the 515+2060~nm [panel (a)] and 2060+515~nm [panel (b)].
We first discuss the 515+2060~nm case shown in Fig.~\ref{fig:ddos_mid_double}(a).
The first short-wavelength pulse induces excitation of electrons from the valence band into the conduction band, as reflected by a positive DDOS in the low-energy region of the conduction band (0--5~eV).
After the second, longer-wavelength pulse, the DDOS shows a redistribution of the excited electron population within the conduction band, with the population in the 0--5~eV range decreasing while that above 5~eV increases.
This redistribution is consistent with additional energy gain of electrons already in the conduction band and may reflect the involvement of intraband dynamics, which are generally expected to be more pronounced at longer wavelengths.
In contrast, for the reverse pulse order (2060+515~nm) shown in Fig.~\ref{fig:ddos_mid_double}(b), the first long-wavelength pulse produces a smaller excited-electron population than in the 515+2060~nm case.
When the second short-wavelength pulse arrives, electrons are excited from the valence band into the conduction band over a similar energy range (0--5~eV).
However, the electron population at energies above 5~eV remains smaller than in the 515+2060~nm sequence.

\begin{figure}[tb]
  \centering
  \includegraphics[width=0.85\linewidth]{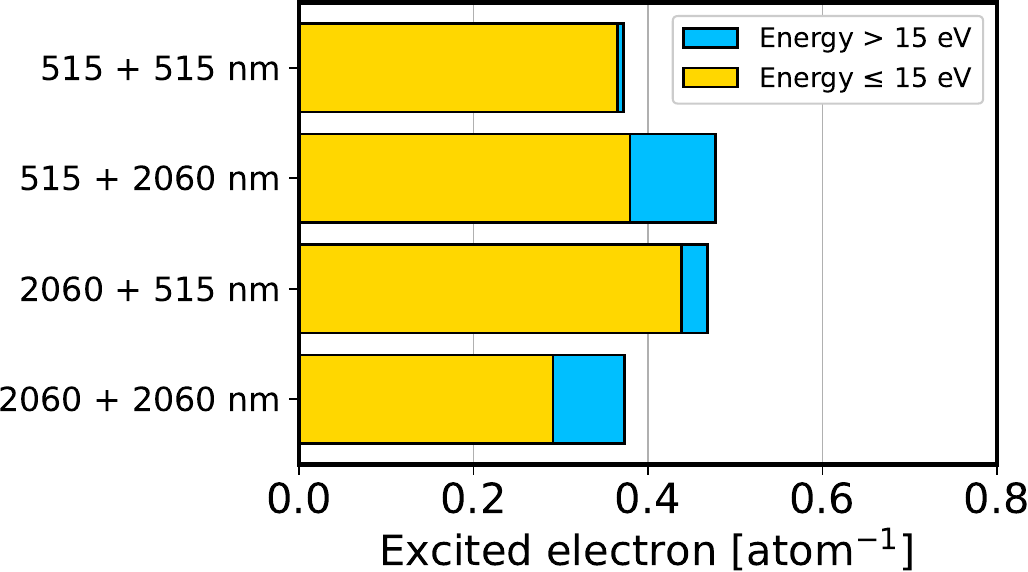}
  \caption{
    Number of electrons excited to states within and beyond the projection energy range $\epsilon_\mathrm{max}=15$~eV after double-pulse irradiation at $I_1=I_2=3.5 \times 10^{12}$~W/cm$^2$.  
  }
  \label{fig:ddos_out_mid_double}
\end{figure}

Figure~\ref{fig:ddos_out_mid_double} compares the populations of electrons excited to energy levels below and above $\epsilon_\mathrm{max}=15$~eV by double-pulse irradiation at $I_1=I_2=3.5\times10^{12}$~W/cm$^2$.
Wavelength combinations involving different colors (515+2060~nm and 2060+515~nm) lead to a larger total number of excited electrons than the monochromatic combinations (515+515~nm and 2060+2060~nm).
Notably, for the two-color pulse sequences 515+2060~nm and 2060+515~nm, the final total number of excited electrons is nearly identical.
However, the population of electrons excited to energies above $\epsilon_\mathrm{max}=15$~eV is larger for the 515+2060~nm.
This indicates that, even though the total number of excited electrons is nearly the same for the two pulse orders, the 515+2060~nm sequence produces a larger number of high-energy electrons, which is consistent with the enhanced energy gain per carrier and may contribute to the increased total energy absorption observed for this pulse sequence.

\subsubsection{\label{subsubsec:delay_time_dependence}Delay-time and relative-phase dependence}

\begin{figure}[tb]
  \centering
  \includegraphics[width=0.9\linewidth]{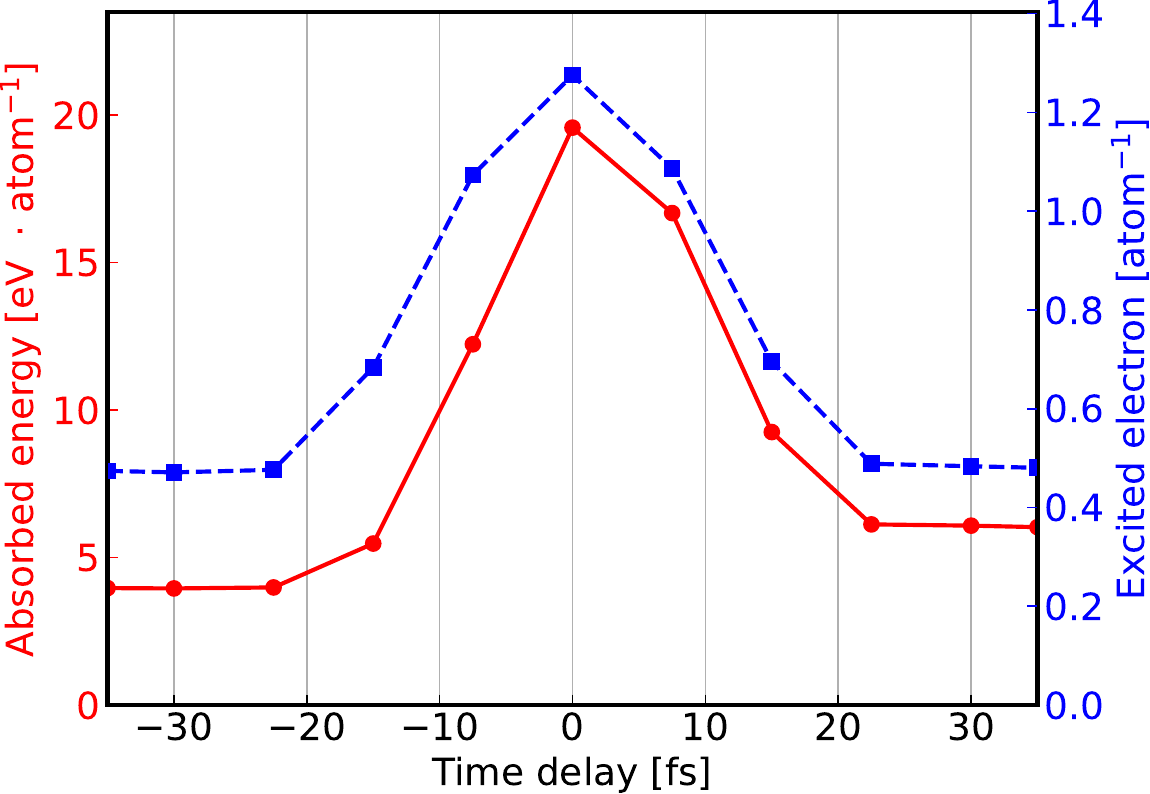}
  \caption{
    Total absorbed energy and number of excited electrons as a function of the delay time $T_{\mathrm{delay}}$ in the 515 + 2060~nm configuration.  
    The intensity is fixed at $I_1 = I_2 = 3.5 \times 10^{12}$~W/cm$^2$.
    A positive delay means that the 515 nm pulse precedes the 2060 nm pulse. 
  }
  \label{fig:delay_dep}
\end{figure}

Figure~\ref{fig:delay_dep} shows the total absorbed energy and the number of excited electrons as functions of the delay time $T_{\mathrm{delay}}$ at $I_1 = I_2 = 3.5 \times 10^{12}$~W/cm$^2$. 
A positive delay means that the 515 nm pulse precedes the 2060 nm pulse. 
When the pulses overlap ($-20 \lesssim T_{\mathrm{delay}} \lesssim 20$), the energy absorption is increased, as we have previously reported \cite{Tani22}. In contrast, when the pulses are well separated, both the absorbed energy and the excited electron number are virtually independent of the delay time.

\begin{figure}[tb]
  \centering
  \includegraphics[width=0.9\linewidth]{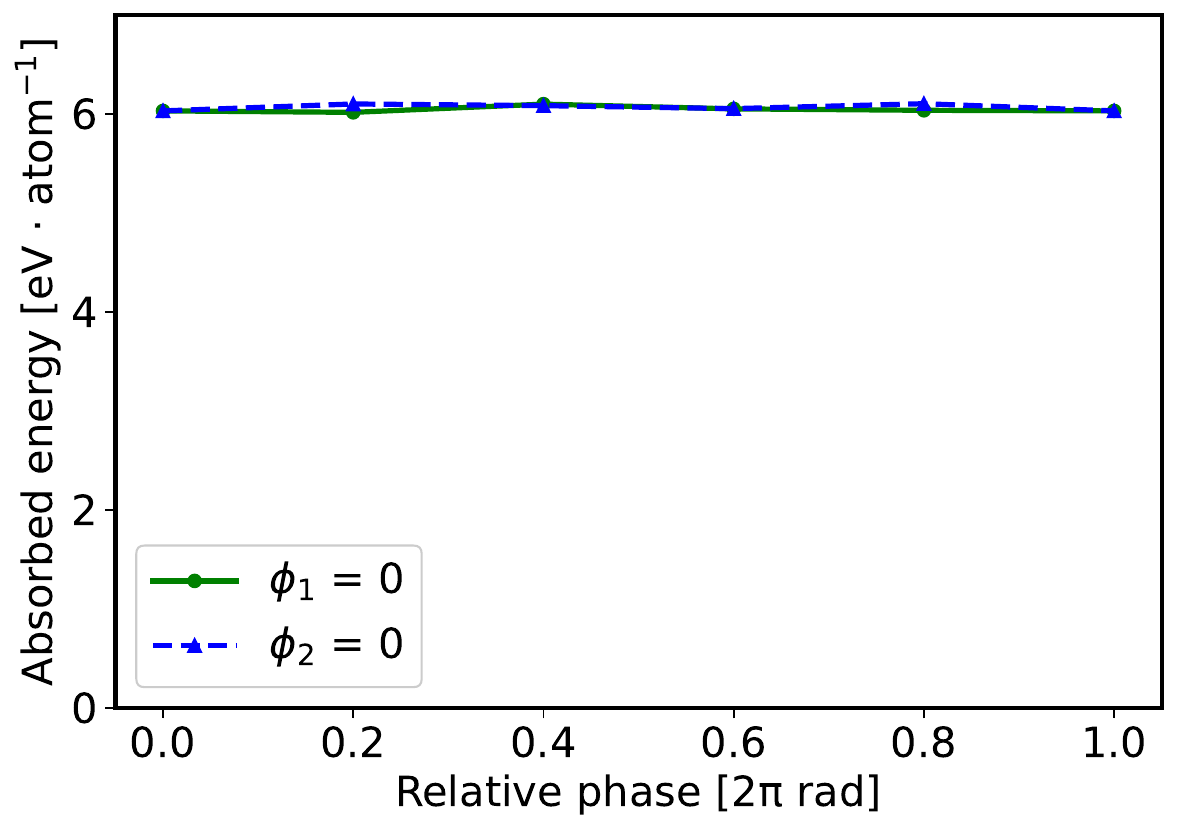}
  \caption{
    Relative phase dependence of the absorbed energy at 515 + 2060~nm, $T_{\mathrm{delay}} = 35$~fs, and $I_1 = I_2 = 3.5 \times 10^{12}$~W/cm$^2$.
    $\phi_1$ and $\phi_2$ denote the carrier–envelope phases of the 515 nm and 2060 nm pulses, respectively.
  }
  \label{fig:phase_dep}
\end{figure}

In Fig.~\ref{fig:phase_dep} we show the dependence of the total absorbed energy on the relative phase $\phi_1 - \phi_2$ for the 515+2060 nm pulse combination at $I_1 = I_2 = 3.5 \times 10^{12}$ W/cm$^2$ and $T_{\mathrm{delay}} = 35$ fs. 
Here, $\phi_1$ and $\phi_2$ denote the carrier–envelope phases of the 515 nm and 2060 nm pulses, respectively.
The absorbed energy remains nearly constant at approximately 6.0 eV/atom, exhibiting negligible dependence on the relative phase.

\subsection{\label{subsec:low}Wavelength-combination dependence at low intensity}

\begin{figure}[tb]
  \centering
  \includegraphics[width=0.9\linewidth]{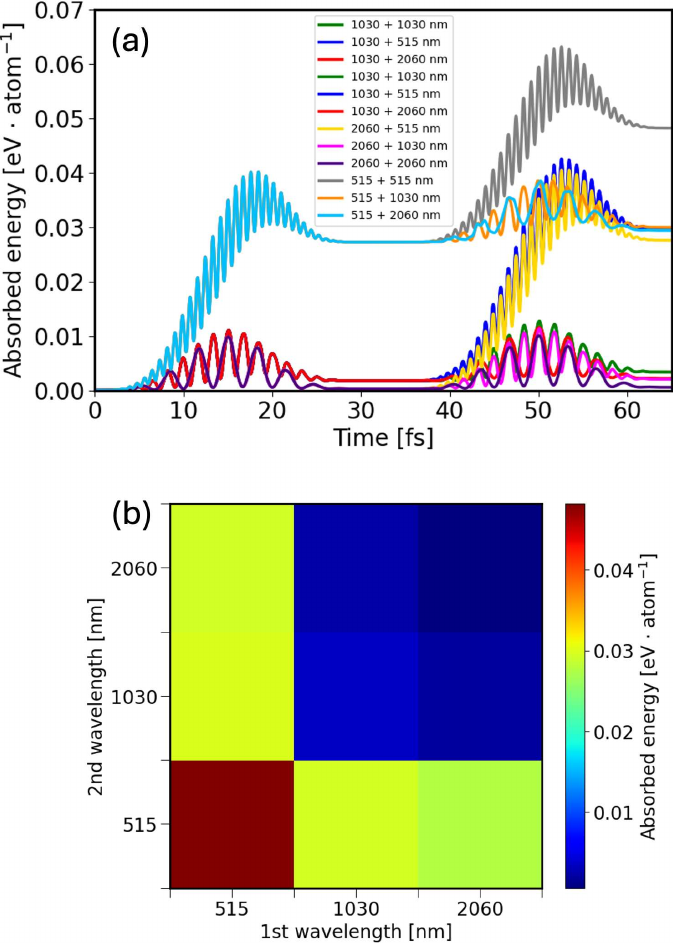}
  \caption{
  (a) Total absorbed energy and (b) number of excited electrons for each wavelength combination with $I_1 = I_2 = 2 \times 10^{11}$~W/cm$^2$ and $T_{\mathrm{delay}} = 35$~fs.}
  \label{fig:abs_low_double}
\end{figure}

\begin{figure}[tb]
  \centering
  \includegraphics[width=0.9\linewidth]{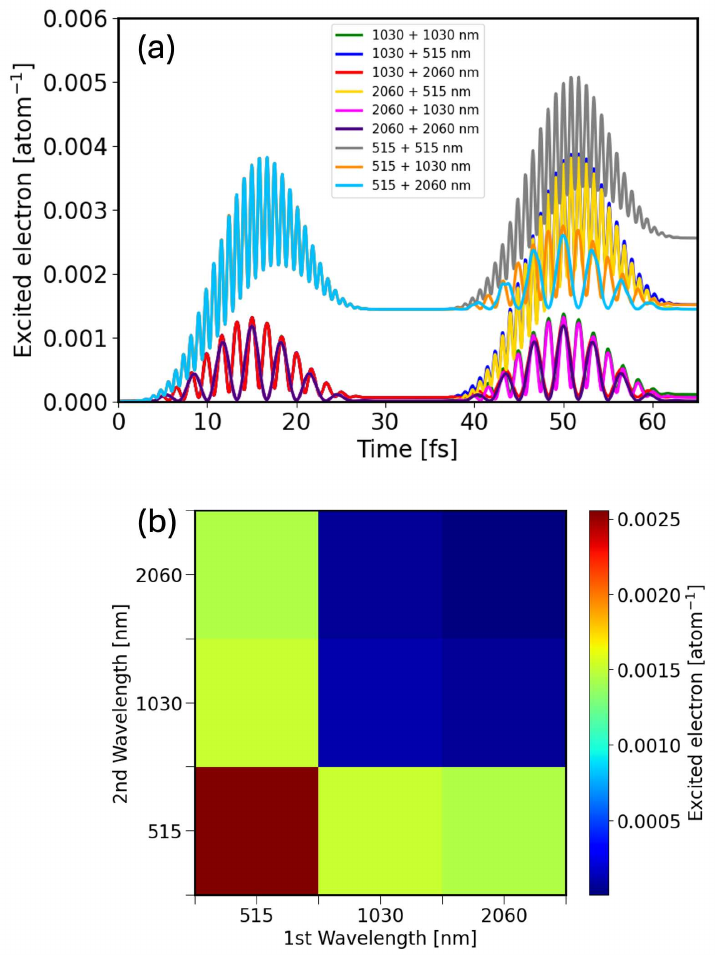}
  \caption{
  (a) Temporal profile of the calculated number of excited carriers and  (b) total number of excited carriers for each wavelength combination with $I_1=I_2=3.5 \times 10^{12}$ W/$\mathrm{cm}^2$ and $T_{\mathrm{delay}}=35$ fs.}
  \label{fig:elc_low_double}
\end{figure}

\begin{figure}[tb]
  \centering
  \includegraphics[width=0.9\linewidth]{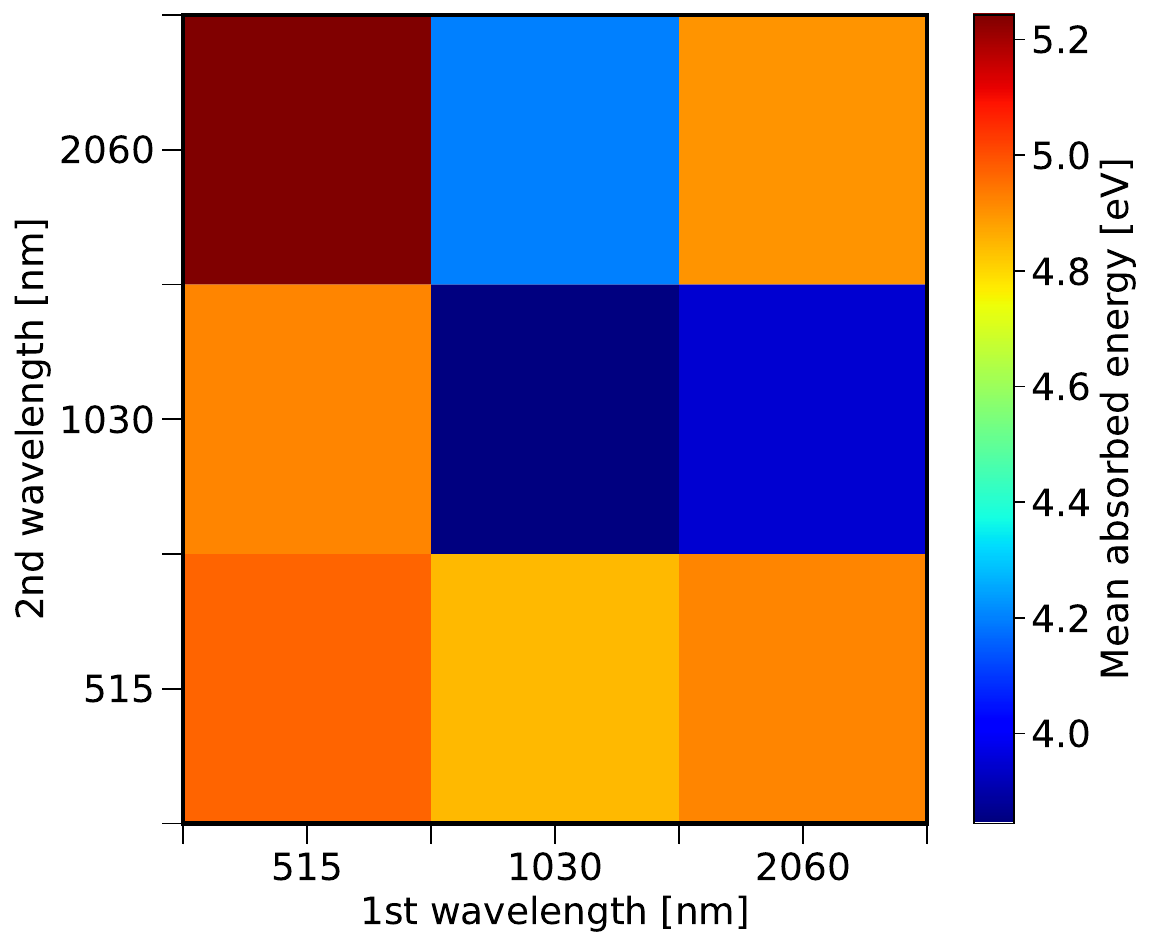}
  \caption{Mean absorbed energy per excited electron for each wavelength combination with $I_1=I_2=2\times 10^{11}$ W/cm$^2$ and $T_{\mathrm{delay}}=35$~fs.}
  \label{fig:mean_low_double}
\end{figure}

In this subsection, we examine a pulse intensity regime lower than that considered in Sec.~\ref{subsec:moderate}, focusing on the case of $I_1 = I_2 = 2 \times 10^{11}$~W/cm$^2$ with $T_{\mathrm{delay}} = 35$~fs.
At this intensity, as shown in Fig.~\ref{fig:abs_low_double}(a), the energy absorption induced by the 515~nm pulse is significantly larger than that induced by the 1030~nm and 2060~nm pulses at $t=35$~fs, immediately after single-pulse irradiation.
Comparing the cases of 515+2060~nm and 2060+515~nm, the final absorbed energy is slightly larger for the former than for the latter, as seen around $t=65$~fs.
As shown in Fig.\ref{fig:abs_low_double}(b), the final absorbed energy is generally higher for wavelength combinations that include the 515 nm pulse, while the dependence on the pulse order for a given wavelength combination is relatively weak.
Figures~\ref{fig:elc_low_double}(a) and (b) show the temporal evolution of the number of excited electrons and their final values, respectively.
The dependence of the excited electron number on the wavelength combination closely follows the trend observed for the absorbed energy.
The mean absorbed energy per carrier, derived from these results, is shown in Fig.~\ref{fig:mean_low_double}, where a clear pulse-order asymmetry is observed, with a significantly higher value for the 515+2060 nm sequence than for the other combinations.

\begin{figure}[tb]
  \centering
  \includegraphics[width=0.9\linewidth]{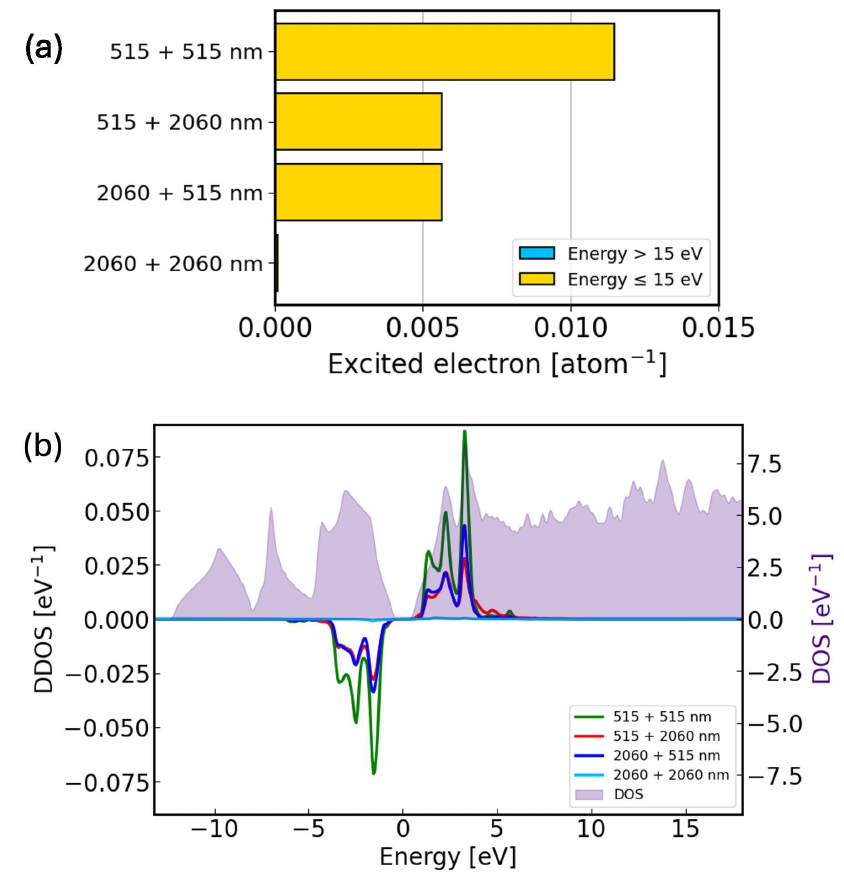}
  \caption{
  (b) Total number of electrons excited to energy levels below and above 15 eV for each wavelength combination and (a) DDOS at the end of the simulation with $I_1=I_2=2 \times 10^{11}$~W/cm$^2$ and $T_{\mathrm{delay}} = 35$~fs.}
  \label{fig:ddos_low_out}
\end{figure}

To elucidate the origin of this behavior, we further analyze the electronic energy distribution by calculating the DDOS at the end of the simulation.
As shown in Fig.\ref{fig:ddos_low_out}(a), the total number of excited electrons is nearly identical for the 515+2060 nm and 2060+515 nm cases.
Furthermore, Fig.\ref{fig:ddos_low_out}(b) shows that the energy distributions of hole generation in the valence band (below 0 eV) are also very similar for the 515+2060 nm (red line) and 2060+515 nm (blue line) sequences.
In contrast, a clear difference appears in the energy distribution of excited electrons in the conduction band (above 0 eV).
Specifically, electrons are more populated below approximately 4 eV for the 2060+515 nm case, whereas electrons at higher energies are populated in the 515+2060 nm sequence.

These results suggest that, even in this relatively low intensity regime, the combination of electron excitation by a short-wavelength pulse followed by additional energy gain by a longer-wavelength pulse enhances the absorbed energy per excited electron.
Nevertheless, because the number of electrons excited by the short-wavelength pulse is much larger than that by the longer-wavelength pulse in this intensity regime, the total absorbed energy is largely governed by the number of excited electrons.

In addition, in the 515+515 nm case in Fig.~\ref{fig:ddos_low_out}(b), a small peak is observed around 5.7 eV.
The energy difference between this peak and the large peak around 3.3 eV is approximately 2.4 eV, which corresponds to the photon energy of the 515 nm pulse (2.41 eV).
This suggests that the feature around 5.7 eV is formed by one-photon excitation by the 515 nm pulse.

\subsection{\label{subsec:high}Wavelength-combination dependence at high intensity}

\begin{figure}[tb]
  \centering
  \includegraphics[width=0.9\linewidth]{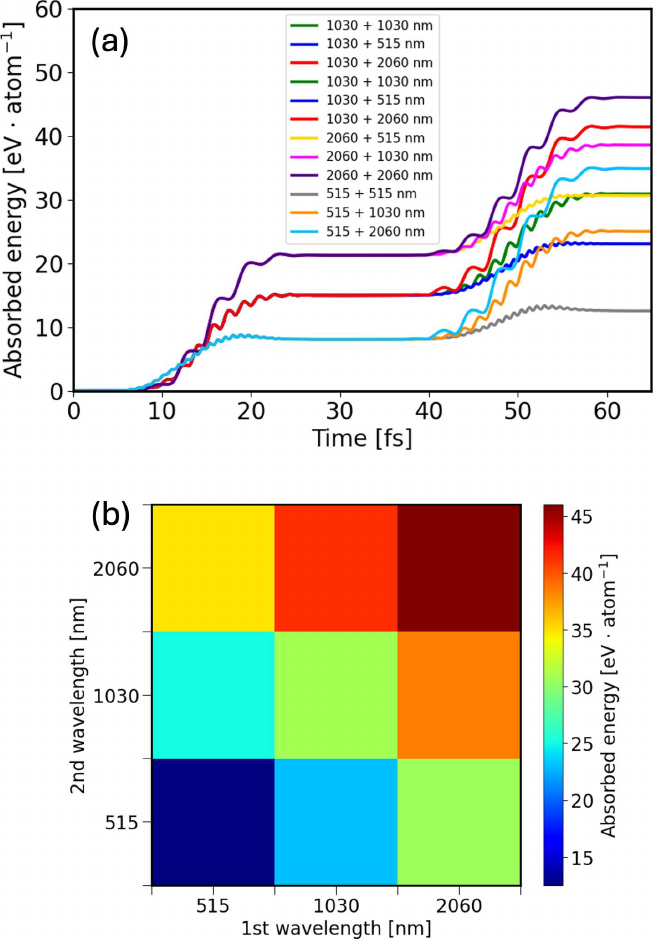}
  \caption{
  (a) Temporal profile of the absorbed energy and (b) total amount of absorbed energy at the end of the simulation for each wavelength combination with $I_1 = I_2 = 10^{13}$ W/cm$^2$ and $T_{\mathrm{delay}} = 35$~fs.}
  \label{fig:abs_high_double}
\end{figure}

Let us now turn to the high-intensity regime.
Here, we consider the case where $I_1 = I_2 = 10^{13}$~W/cm$^2$ and $T_{\mathrm{delay}} = 35$~fs.
Figure~\ref{fig:abs_high_double}(a) shows the temporal evolution of the absorbed energy for each wavelength combination.
Immediately after both the first and second pulses ($t=35,~65$~fs), configurations involving longer wavelengths exhibit significantly larger energy transfer. 
As shown in Fig.~\ref{fig:abs_high_double}(b), the total absorbed energy tends to be larger when the pulse sequence includes the longer wavelength component of 2060nm.
Furthermore, when comparing the pulse order for a given pair of wavelengths, the sequence in which the shorter-wavelength pulse precedes the longer-wavelength pulse consistently results in greater energy absorption than the reverse order.
Specifically, the following relations hold: 515+1030 nm $>$ 1030+515nm, 515+2060nm $>$ 2060+515nm, and 1030+2060nm $>$ 2060+1030nm.

\begin{figure}[tb]
  \centering
  \includegraphics[width=0.9\linewidth]{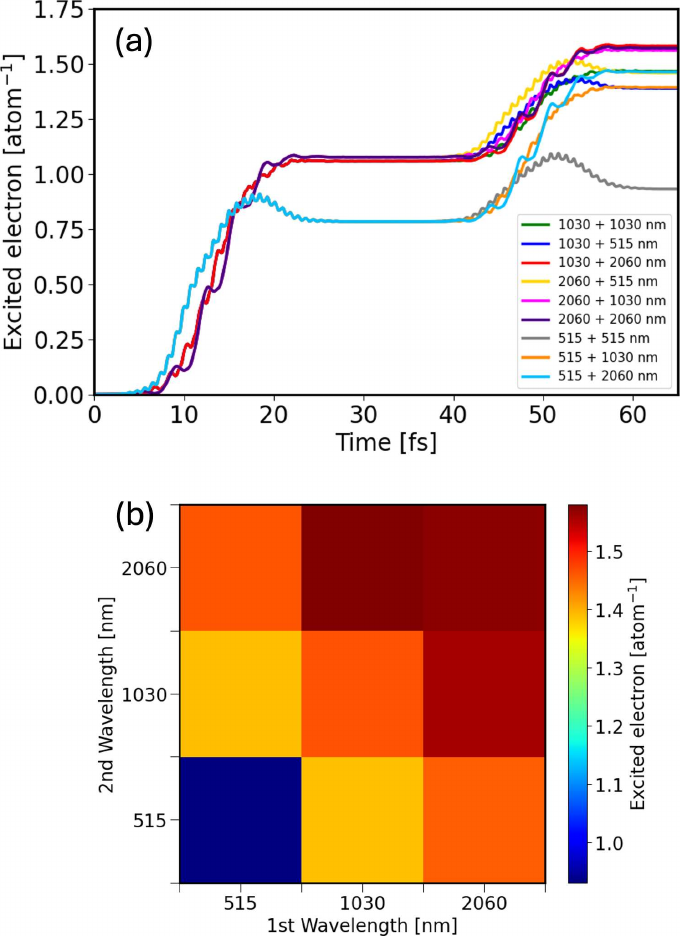}
  \caption{
  (a) Temporal profile and (b) final number of excited electrons for each wavelength combination with $I_1=I_2= 10^{13}$ W/$\mathrm{cm}^2$ and $T_{\mathrm{delay}}=35$ fs.}
  \label{fig:elc_high_double}
\end{figure}

Figure~\ref{fig:elc_high_double}(a) shows the temporal evolution of the number of excited electrons. 
Excitation induced by both the first and second pulses is larger for longer wavelengths, and in particular, the 2060 nm pulse exhibits a step-like increase in the excited-electron number, suggesting the contribution of tunneling ionization. 
As shown in Fig.~\ref{fig:elc_high_double}(b), reversing the pulse order for a given wavelength combination results in nearly identical final numbers of excited electrons. 
However, in the 2060+2060 nm case, the increase in excited-electron number induced by the second pulse is smaller than that induced by the first pulse.

\begin{figure}[tb]
  \centering
  \includegraphics[width=0.9\linewidth]{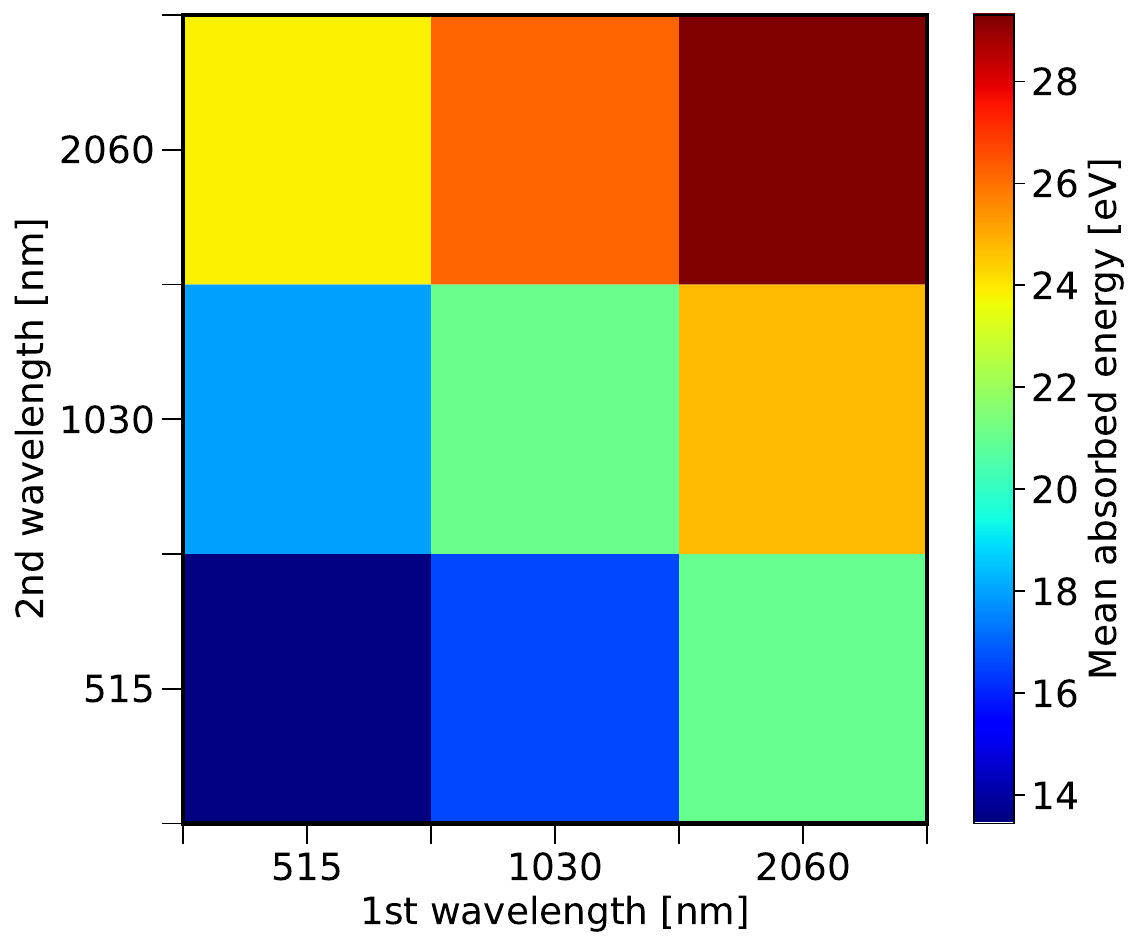}
  \caption{Mean absorbed energy per excited electron for each wavelength combination with $I_1=I_2=10^{13}$ W/cm$^2$ and $T_{\mathrm{delay}}=35$~fs.}
  \label{fig:mean_high_double}
\end{figure}

Furthermore, for the three combinations 2060+2060 nm, 2060+1030 nm, and 1030+2060 nm, the temporal evolution of the excited-electron number is nearly identical, whereas the temporal evolution of the absorbed energy differs significantly, as shown in Fig.~\ref{fig:abs_high_double}(a). 
The final absorbed energy follows the order 2060+2060 nm $>$ 1030+2060 nm $>$ 2060+1030 nm. 
A similar trend is observed for the mean absorbed energy per excited electron shown in Fig.~\ref{fig:mean_high_double}, indicating that the differences in absorbed energy originate primarily from variations in the energy gain per electron rather than from the total number of excited electrons.

Focusing on the high-intensity 2060+2060 nm case, Fig.~\ref{fig:abs_high_double}(a) shows that the energy increments induced by the first and second pulses are comparable, whereas Fig.~\ref{fig:elc_high_double}(a) indicates that the increase in excited-electron number due to the second pulse is smaller than that due to the first pulse. 
This clearly demonstrates that the absorbed energy is not proportional to the number of excited electrons.

\begin{figure}[tb]
  \centering
  \includegraphics[width=0.95\linewidth]{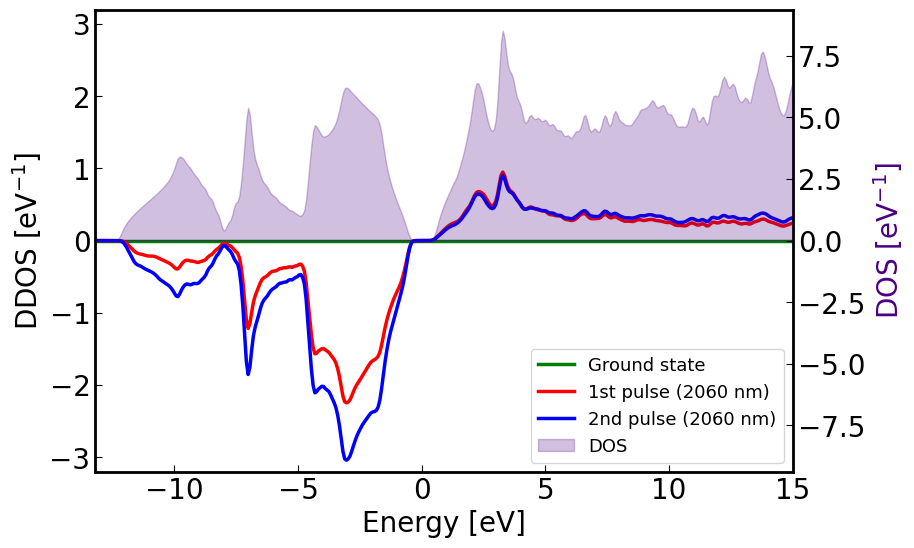}
  \caption{
  DDOS at $t=35$ and $65$~fs, corresponding to the times immediately after the first and second laser pulses, respectively, for $I_1=I_2=10^{13}$~W/cm$^2$ for the 2060+2060~nm.}
  \label{fig:ddos_high_double}
\end{figure}

To clarify the underlying mechanism, we compare the DDOS immediately after the first and second pulses for the 2060+2060 nm configuration as shown in Fig.~\ref{fig:ddos_high_double}. 
After the first pulse, excitation occurs from a broad energy range within the valence band. 
A similar energy region contributes to excitation after the second pulse, however, the number of newly generated excited electrons is smaller.
Despite this, the absorbed energy remains comparable, indicating that the second pulse does not primarily enhance valence-to-conduction excitation but instead provides additional energy to electrons already promoted to the conduction band.
In other words, intraband acceleration and redistribution toward higher-energy states within the conduction band likely dominate during the second pulse. Indeed, as shown in Fig.~\ref{fig:ddos_high_double}, the energy distribution of conduction-band electrons immediately after the first and second pulses does not change significantly within the projection window, implying that the second pulse accelerates electrons to energies beyond the projection range ($\epsilon_\mathrm{max}=15$~eV).

\begin{figure}[tb]
  \centering
  \includegraphics[width=0.95\linewidth]{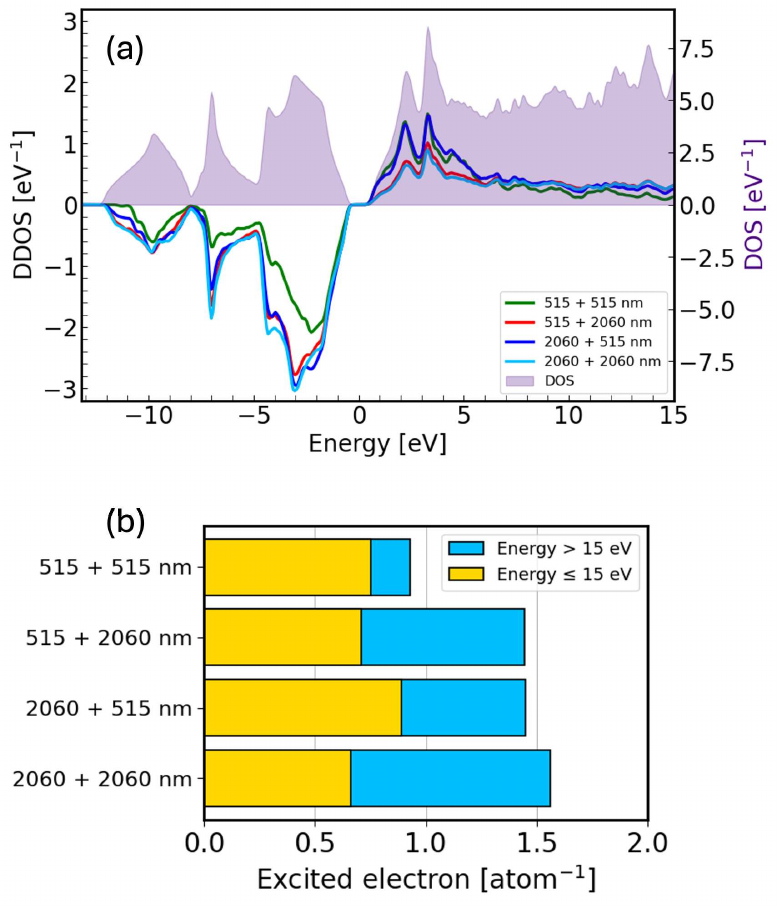}
  \caption{
  (a) DDOS at the end of the simulation and (b) total number of electrons excited to energy levels below and above 15 eV for each wavelength combination with $I_1=I_2=10^{13}$~W/cm$^2$ and $T_{\mathrm{delay}} = 35$~fs.}
 \label{fig:ddos_high_out}
\end{figure}

Figure~\ref{fig:ddos_high_out}(a) shows the DDOS after double-pulse irradiation for each wavelength combination.
Focusing first on the valence band, electron excitation is observed over a wide energy range for all wavelength combinations. 
The total number of excited electrons is markedly smaller for the 515+515~nm case than for the other combinations.
Turning to the conduction band distribution, in the relatively low-energy region (0–approximately 5eV), similar distributions are observed when the second pulse is 515nm (515+515nm and 2060+515nm), while another similar pair appears when the second pulse is 2060nm (515+2060nm and 2060+2060nm). 
However, the occupation in this low-energy region is relatively smaller when the second pulse is 2060nm.
Despite this reduction at low energies, the total number of excited electrons in the 515+2060nm and 2060+2060nm cases is comparable to or larger than that in the 2060+515nm case. This suggests that when the second pulse is 2060nm, electrons are redistributed from the low-energy conduction-band region to higher-energy states.

To quantify this redistribution, Fig.\ref{fig:ddos_high_out}(b) separates the excited electrons into contributions below and above 15eV. For the 2060+2060nm configuration, the total number of excited electrons is the largest, and more than half of them occupy high-energy states above 15eV. Furthermore, although the total number of excited electrons is nearly the same for 515+2060nm and 2060+515nm, the former exhibits a larger population in the high-energy region above 15~eV.
These results indicate that in configurations involving long-wavelength pulses, the enhanced absorption cannot be attributed solely to increased interband excitation. Instead, strong intraband acceleration driven by the intense long-wavelength field plays a significant role, promoting electron redistribution toward higher-energy states within the conduction band.

The relaxation process of excited electrons, which is not considered in our TDDFT framework, can be roughly estimated using the impact ionization rate (IIR) investigated via first-principles calculations~\cite{Kotani10}.
Kotani and van Schilfgaarde~\cite{Kotani10} have reported the IIR for excited electrons with energies of $1\sim5$ eV, measured from the conduction band minimum.
In the present study, the first pulse with a peak intensity of $10^{13}$~W/cm$^2$ excites approximately 1.1 electrons/atom (Fig.~\ref{fig:elc_high_double}(a)), and the peak of the excited electron energy distribution remains below 5 eV (Fig.~\ref{fig:abs_high_double}). 
By combining these observations with the results of Ref.~\cite{Kotani10}, we estimate that the number of impact ionization events during the 35 fs double-pulse delay is at most 3.85 events/atom. 
These estimates suggest that, for higher-intensity pulses and longer delays, energetic relaxation of the initially excited electrons may become non-negligible. 
Therefore, the corresponding results should be interpreted with some care, although they still provide useful insight into the early-stage electronic response within our TDDFT framework.

\subsection{\label{subsec:intensity}Intensity dependence}

\begin{figure}[tb]
  \centering
  \includegraphics[width=0.9\linewidth]{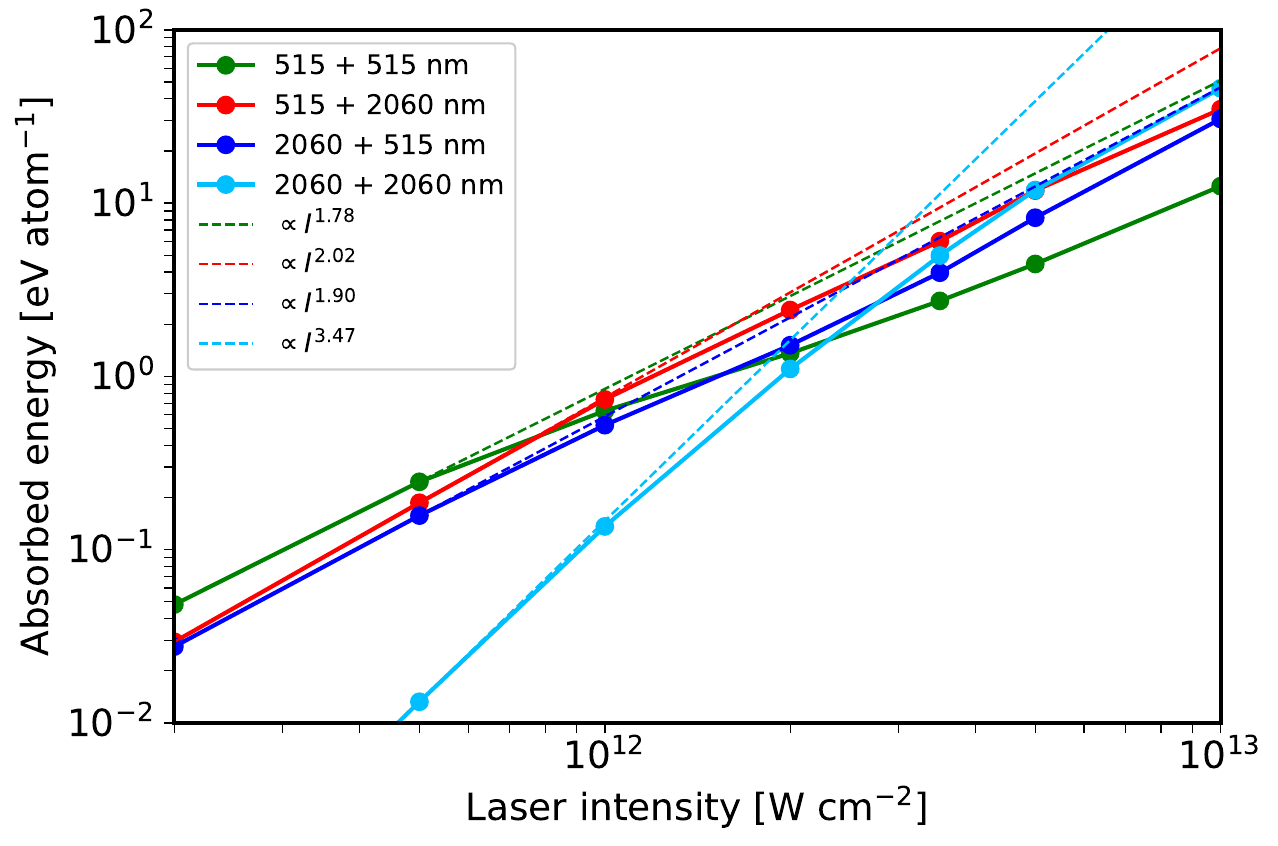}
  \caption{
  Intensity dependence of total absorbed energy for selected wavelength combinations: 515+515~nm, 515+2060~nm, 2060+515~nm, and 2060+2060~nm. 
  The time delay is fixed at $T_{\mathrm{delay}} = 35$~fs.
  The dashed lines represent power-law fits $W \propto I^{s}$ obtained from the two lowest-intensity points.
  }
  \label{fig:intensity_scan}
\end{figure}

We now examine how the total absorbed energy varies with peak intensity for each of the four wavelength combinations: 515+515 nm, 515+2060 nm, 2060+515 nm, and 2060+2060 nm.
Figure~\ref{fig:intensity_scan} shows the total absorbed energy $\mathrm{W_{tot}}$ as a function of peak intensity $I$ for $T_{\mathrm{delay}} = 35$~fs.
The dashed lines represent power-law fits $\mathrm{W_{tot}} \propto I^{s}$ obtained from the two lowest-intensity points ($2\times10^{11}$ and $5\times10^{11}$ W/cm$^2$). 
The Keldysh parameter $\gamma$ evaluated with a direct band gap of 3.1 eV for Si [see Fig.~\ref{fig:Re_Im}(b)] is, for 515 nm, $\gamma > 1$ throughout the present intensity range, corresponding to the multiphoton regime, and, for 2060 nm, $\gamma \sim 1$ at $I \sim 3.16 \times 10^{12}\,{\rm W/cm^2}$, indicating a crossover toward the tunneling regime.

For the wavelength combinations including 515 nm with a 2.41 eV photon energy, the low-intensity scaling is close to quadratic, consistent with two-photon interband excitation.
For the 2060+2060 nm case, the exponent (3.47) is significantly smaller than the value of 6 expected from the photon energy. This discrepancy can be attributed to the spectral profile of the actual pulse. The electric field used in this study is not Gaussian, and its Fourier spectrum includes components with energies higher than the central value of 0.601 eV. In particular, a component around 0.89 eV appears with a relative spectral power of approximately $10^{-2}$ with respect to the peak. Such higher-energy components can promote three- or four-photon excitation, thereby reducing the apparent multiphoton order inferred from the central wavelength alone.
As the intensity increases, the data deviate downward from the low-intensity power-law scaling, as commonly observed in strong-field excitation.

For 2060+515 nm, the exponent is slightly below two, but its high-intensity behavior differs from the other cases. 
While the low-intensity scaling is governed by interband excitation induced by the 515 nm component, the contribution of field-driven excitation associated with the 2060 nm component becomes more significant at higher intensities, leading to a distinct deviation from the simple power-law behavior.

\section{\label{sec:conclusion}CONCLUSION}

In this study, we have systematically investigated ultrafast energy absorption in crystalline Si under femtosecond double-pulse laser irradiation using TDDFT, focusing on the roles of wavelength combination and peak intensity.

We have found that the optimal wavelength combination for energy absorption strongly depends on the intensity regime.
At low intensities, absorption is governed by multiphoton interband excitation and is enhanced for wavelength combinations including the shorter wavelength (515 nm).
In contrast, at high intensities, field-driven excitation and intraband acceleration induced by the longer wavelength (2060 nm) become dominant, leading to enhanced energy absorption. 

In the intermediate regime, the 515+2060 nm configuration yields the largest absorption, indicating an efficient interplay between carrier generation by the short-wavelength pulse and subsequent intraband acceleration driven by the long-wavelength pulse.
A pronounced pulse-order effect is also observed, with higher absorption when the short-wavelength pulse precedes the long-wavelength pulse, highlighting the importance of the nonequilibrium electronic energy distribution created by the first pulse.

These results demonstrate that ultrafast energy deposition can be effectively controlled by tailoring the wavelength combination, intensity, and the pulse sequence. 
Our findings provide a microscopic understanding of two-color laser-matter interaction and offer useful guidelines for optimizing ultrafast laser processing and optical control in semiconductors and wide-bandgap materials.

\begin{acknowledgments}
This research was supported by MEXT Quantum Leap Flagship Program (MEXT Q-LEAP) Grant No. JPMXS0118067246, JSPS KAKENHI Grants No. JP24H00427, No. JP24K01224, No. JP24K23024, No. JP25K17367, No. 25H00704, and JST SPRING, Grant Number JPMJSP2108.
This work was also partially supported by JST K Program Grant Number JPMJKP24M1, Japan and the RIKEN TRIP initiative.
E.G. gratefully acknowledges support from World-leading Innovative Graduate Study Program Co-designing Future Society (WINGS-CFS) of the University of Tokyo.
The numerical calculations are partially performed on supercomputer Oakbridge-CX and Wisteria/BDEC-01 (the University of Tokyo).
\end{acknowledgments}

\appendix
\section{Ground state and linear response}
\label{sec:appendix}
The density of states (DOS) at the ground state is calculated using the following expression:
\begin{equation}
    f(\epsilon) = \frac{2}{N_k} \sum_{\mathbf{k}}\sum_{b} \delta(\epsilon - E_{b,\mathbf{k}}),
\end{equation}
where $N_k$ is the total number of $k$-points, $E_{b,\mathbf{k}}$ denotes the energy level at band index $b$ and wave vector $\mathbf{k}$, and $\delta$ is the Dirac delta function.  
In practical implementation, the delta function is replaced by a Gaussian broadening to yield a smooth spectrum. 

\begin{figure}[tb]
  \centering
  \includegraphics[width=0.9\linewidth]{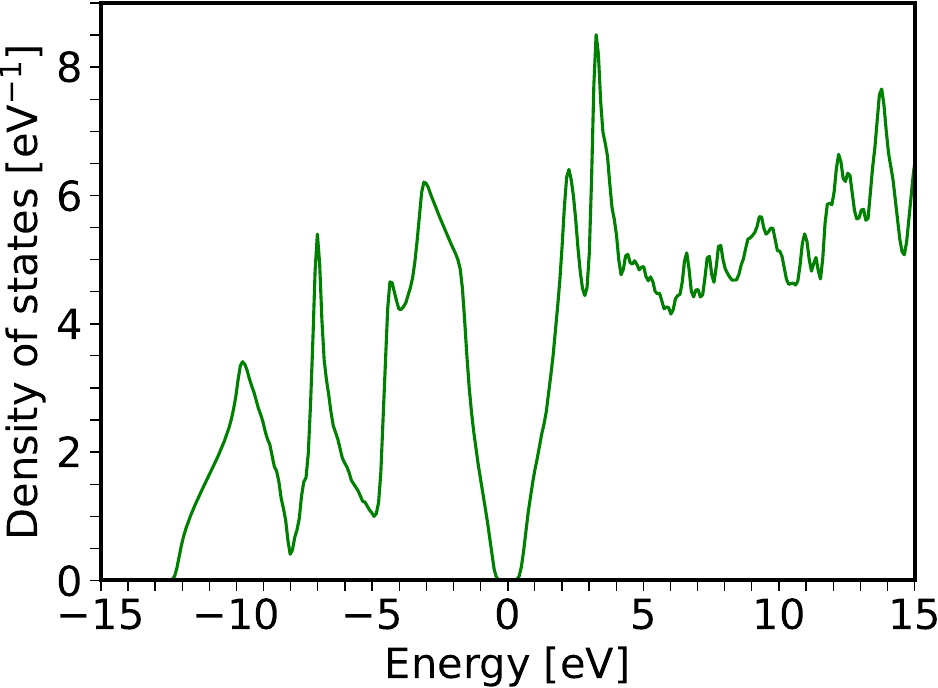}
  \caption{The calculated density of states of Si.
  0 eV is set to the Fermi level.}
  \label{dos}
\end{figure}

\begin{figure}[t]
  \centering
  \includegraphics[width=0.85\linewidth]{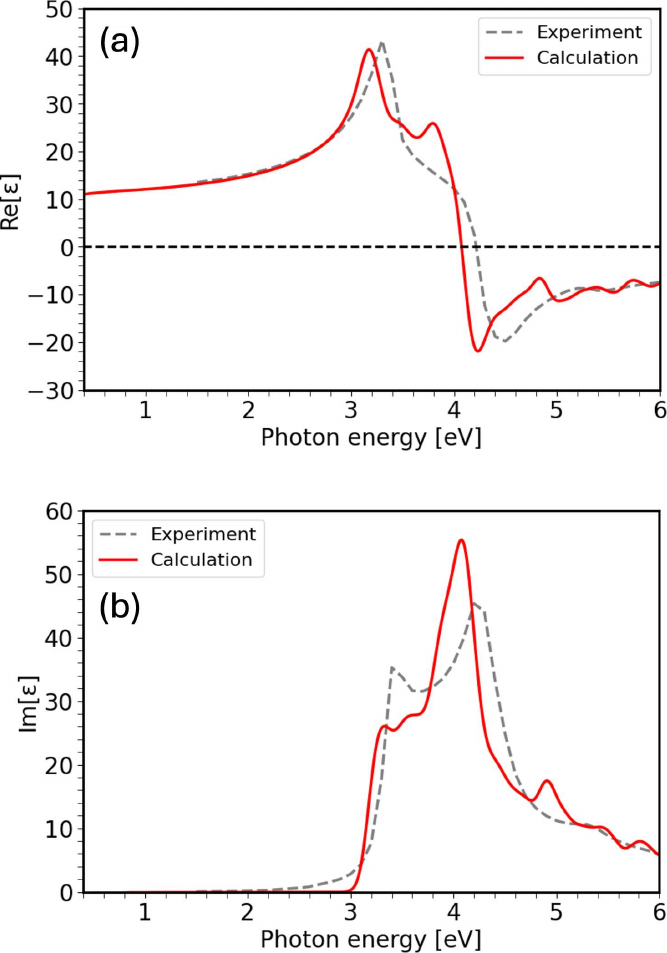}
  \caption{
    The calculated dielectric function of crystalline Si.  
    The red solid line represents (a) the real and (b) the imaginary part from our simulation; the gray dashed line shows the experimental reference~\cite{Schinke15}. 
  }
  \label{fig:Re_Im}
\end{figure}

Figure~\ref{dos} shows the calculated density of states of crystalline silicon.
The direct bandgap energy is estimated from the imaginary part of the dielectric function, obtained via time propagation after an impulsive momentum kick.  
This perturbation is realized by a step-like vector potential,
\begin{equation}
    \mathbf{A}(t) = \left\{
        \begin{array}{ll}
            -\mathbf{A}_0 & (t \geq 0)\\
            0 & (t < 0)
        \end{array}
    \right.,
    \label{A0}
\end{equation}
which corresponds to an impulsive electric field,
\begin{equation}
    \mathbf{E}(t) = \mathbf{A}_0 \delta(t).
    \label{E0}
\end{equation}
Since this electric field has a flat power spectrum over all frequencies, the optical conductivity in the $m$-direction is calculated from the temporal Fourier transform of the current response as:
\begin{equation}
    \sigma_m(\omega) = \frac{\hat{J}_m(\omega)}{A_{0m}} \quad (m = x, y, z),
\end{equation}
where $\hat{J}_m(\omega)$ is the $m$-component of the Fourier-transformed current density.
Assuming an isotropic medium, the dielectric function is then given by:
\begin{equation}
    \varepsilon_m(\omega) = 1 + 4\pi i \frac{\sigma_m(\omega)}{\omega}.
\end{equation}
The imaginary part of the calculated dielectric function agrees well with experimental results~\cite{Schinke15}, as shown in Fig.~\ref{fig:Re_Im}.
The estimated direct bandgap energy is 3.1 eV, which is in excellent agreement with the experimental value of 3.2 eV~\cite{Aspnes83}.  
Note that the direct bandgap energy, rather than the indirect one (see Fig.~\ref{dos}), is relevant for optical absorption in our calculations.
Overall, the ground state obtained here accurately reproduces the electronic properties of real crystalline silicon and serves as a reliable basis for time-dependent simulations.

\bibliography{ref}

\end{document}